\documentclass[prc,showpacs,twocolumn]{revtex4}
\usepackage{graphicx}
\usepackage{amsmath,amssymb}
\usepackage{mathrsfs}
\def\pt{$p_T$}
\def\qqb{$q\bar q$}
\def\dis{distribution}
\def\ep{$(\eta,\phi)$}
\def\qaq{$q$ and $\bar q$}
\def\bq{\begin{eqnarray}}
\def\eq{\end{eqnarray}}
\def\fq{$F_q(M)$}
\def\fl{fluctuation}
\begin{document}

\title{\Large {\bf Local Multiplicity Fluctuations as a Signature of Critical Hadronization at LHC}}
\author
 {\bf Rudolph C. Hwa$^1$ and C.\ B.\ Yang$^{1,2}$}
\affiliation
{$^1$Institute of Theoretical Science and Department of
Physics\\ University of Oregon, Eugene, OR 97403-5203, USA\\
$^2$Institute of Particle Physics, Central China Normal
University, Wuhan 430079, P.\ R.\ China}
\date{\today}

\begin{abstract}

In central Pb-Pb collisions at LHC the multiplicity of particles produced is so high that it should become feasible to examine the nature of transition from the deconfined quark-gluon state to the confined hadron state by methods that rely on the availability of high multiplicity events. We consider four classes of the transition process ranging from critical behavior to totally random behavior, depending on whether or not  there is clustering of quarks and on whether or not there is contraction of dense clusters due to confinement. Fluctuations of bin multiplicities in each event are quantified, and then the event-by-event fluctuations of spatial patterns are analyzed.  A sequence of measures are proposed and are shown to be effective in capturing the essence of the differences among the classes of simulated events. It is demonstrated that a specific index has a low value for critical transition but a larger value if the hadronization process is random.

\end{abstract}

\maketitle
\section{Introduction}

Phase transition has always been a subject of great interest in many fields. The possibility of observing evidences for the critical point in heavy-ion collisions has invigorated extensive experimental programs at various laboratories \cite{mb,bm}. The physics of QCD critical point concerns a dense system of strongly-interacting matter that is in thermal and chemical equilibrium and is at the end of the phase boundary between quark and hadron phases \cite{ms}. The search for signals of that boundary involves experiments at energies where high baryon density can be produced. At the Large Hadron Collider (LHC) where the collision energy is much higher, the physics of hadronization involves issues that are different from that probed at lower energies where high chemical potential is expected. Since a hot and dense plasma is produced at LHC, the deconfined state persistes for a long time before the system is dilute enough to undergo transition to the hadron phase. That transition may or may not be recognizable as a critical phenomenon, since hadronization takes place on the surface over a period of time while the system expands. The accumulation of hadrons emitted over that period can smear out any signal of interest even in the best circumstance for critical transition, which is not assured on theoretical grounds. Nevertheless, or perhaps particularly because of the difficulty in detecting revealing signals, it is of interest to investigate whether appropriate measures exist.   It is the aim of this work to find the most effective way to extract dynamical signals indicative of collective behavior of a system transitioning from quarks to hadrons. Our search does not rely on the validity of any theoretical view on the possibility of critical behavior
 in heavy-ion collisions at very high energy.
 
Despite the fact that a large amount of data on Pb-Pb collisions at LHC has been produced in the past year \cite{fa}, so far no spectacular departure from earlier expectations \cite{na} by extrapolation from the Relativistic Heavy-Ion Collider (RHIC) has been reported. It is not certain whether it is because there is no unexpected new physics, or because unanticipated physics is not revealed in the conventional observables. We venture here to ask whether there may be an area of investigation of the LHC data that has thus far been neglected but may prove to be fruitful when suitable measures are used.

An obviously outstanding feature of the data obtained at LHC is that the total multiplicity $N_{\rm ch}$ of charged particles produced is unprecedentedly high, around $6\times 10^3$ in central collisions at 2.76 TeV \cite{ka}. Observables that rely on large  $N_{\rm ch}$ can thus be exploited in ways that could not have been possible at lower energies, thereby forming a frontier that has not been explored. Another aspect of central Pb-Pb collisions is that  there is surely a deconfined system of quarks and gluons, which must undergo some kind of transition to the confined state of hadrons. Theoretically, the use of Cooper-Frye scheme \cite{cf} to calculate the hadronic properties avoids the issue about the nature of the deconfinement-to-confinement transition. In our investigation we leave open the question of what the nature of that transition is. We propose four scenarios that range from critical behavior on one end to non-critical on the other. The focus is on the discovery of measures that can possibly distinguish those cases. To test whether the  proposed measures are feasible experimentally, we simulate the events in each case and examine the effectiveness of the measures.

The theme of our investigation is about the fluctuations of spatial patterns during the quark-hadron transition. The momentum of each particle is usually expressed in terms of $(p_T, \eta, \phi)$, where $p_T$ is the transverse momentum, $\eta$ the pseudo-rapidity, and $\phi$ the azimuthal angle. For any interval of $p_T$ one can study the $(\eta,\phi)$ distribution in any given event, called the lego plot when shown in some finite-size binning. For convenience, we refer to that distribution as a spatial pattern, which changes from event to event. The basic question is whether the fluctuations of those patterns contain any information about the nature of the quark-hadron transition. To study those patterns one needs good resolution from the experimental side and efficient description from the theoretical side. Furthermore, one should not integrate over all $p_T$ because the superposition of different patterns at different $\Delta p_T$ intervals can smear out all recognizable features. Thus to be able to have high resolution in all $(p_T, \eta, \phi)$ variables as well as to have enough particles in small bins in each variable requires high event multiplicities. That is where the LHC data become extremely useful.

There are recent reports from ALICE on the event-by-event fluctuations of global observables in Pb-Pb collisions \cite{sh,pc}. Non-statistical \fl s of mean $p_T$ are found to be lower than a scaling behavior in event multiplicity as the collisions become more central. No explanation of the phenomenon is given in terms of known models. Our study here is on local multiplicity \fl s in precisely those central collisions.

There are two kinds of \fl s that we shall consider. One is the \fl\ of bin-multiplicity from bin to bin in an event, and the other is the event-to-event \fl\ of the spatial patterns. On the former it can be just random \fl s, but it can also be in the form of clusters of all sizes, as in second-order phase-transition \cite{jjb}. Both types of patterns will be generated to initiate quark configurations in the $(\eta,\phi)$ space before and during time evolution throughout the period of hadronization for two distinct classes of models. We shall construct an algorithm for simulating the global effect of color confinement through contractions of dense clusters between time steps in the hadronization period and the opposite effect of thermal agitation by randomization right after each contraction. Pion formation according to chosen criteria can occur throughout the process, resulting in an event \dis\ of pions in $(p_T,\eta,\phi)$. That \dis\ is then analyzed by factorial moments to filter out statistical \fl s. Repeating the simulation over many times generates event-by-event \fl s which turn out to be crucial to the finding of revealing signatures. Large deviations from the average event structure are possible but very rare; however, they can make significant influence on the measure that we shall propose.

\section{Local Multiplicity Fluctuations with Critical Clustering}

In search for ways to simulate configurations that have a wide range of characteristics without theoretical prejudices, ranging from critical to non-critical cases, we focus in this section on finding a simple procedure to generate configurations that correspond to critical behavior. We first
 review  a number of related investigations on the local behaviors of multiplicity fluctuations of systems undergoing a second-order phase transition describable by the Ginzburg-Landau (GL) theory.  They are then to be connected to a distribution of cluster production that 
 can readily be used  to simulate initial configurations before hadronization begins. Only one parameter is needed to describe the clustering. Later, for the non-critical case it is only necessary to change that parameter so as to obtain random \dis s.

It was suggested that the normalized factorial moments $F_q$ can be used as a quantitative measure of local fluctuations \cite{bp}.  $F_q$ is defined by
\begin{eqnarray}
F_q(\delta) = {\left<n!/(n-q)!\right>\over \left<n\right>^q},     \label{2.1}
\end{eqnarray}
where $n$ is the multiplicity in a bin of size $\delta^d$ in a d-dimensional phase space, and the averages are performed over many events.  A power-law behavior
\begin{eqnarray}
F_q(\delta) \propto \delta^{-\varphi_q}     \label{2.2}
\end{eqnarray}
over a range of small $\delta$ is referred to as intermittency, and has been observed in many systems of collisions \cite{kd}.  The virtue of $F_q$ is that it filters out statistical fluctuations, as can be seen as follows.  If the multiplicity distribution $P_n$ can be written as a convolution of the Poisson distribution $P_{\bar n}^0(n)$ and a dynamical distribution $D(m)$, i.e.,
\begin{eqnarray}
P_n = \int_0^{\infty} dm{m^n\over n!}e^{-m}D(m),     \label{2.3}
\end{eqnarray}
then the numerator of $F_q$ is
\begin{eqnarray}
\sum_{n=q}^{\infty} {n!\over (n-q)!} P_n = \int_0^{\infty} dmm^qD(m),     \label{2.4}
\end{eqnarray}
which is a simple moment of $D(m)$.  Thus, if the dynamics is trivial, i.e., $D(m)= \delta(m-\bar n)$, then $F_q = 1$ for all $q$.  Any deviation from 1 is then a measure of non-trivial dynamical fluctuations, and a power-law behavior in Eq.\ (\ref{2.2}) would suggest a dynamics that is not characterized by a particular scale \cite{bp}.  The footprint of a PT that has fluctuations of all scales may then be possibly observed in the measurement of intermittency \cite{bh}.

To obtain a theoretical quantification of PT in terms of $F_q$, a study of second-order PT in the Ginzburg-Landau theory was carried out in Ref.\ \cite{hn}, in which the order parameter is identified with the multiplicity density.  It is found that to a high degree of accuracy $F_q$ satisfies the power-law behavior
\begin{eqnarray}
F_q \propto F_2^{\beta_q},     \label{2.5}
\end{eqnarray}
where
\begin{eqnarray}
\beta_q = (q-1)^{\nu}, \qquad\quad \nu = 1.304,     \label{2.6}
\end{eqnarray}
essentially independent of the details of the GL parameters.  Such a behavior was experimentally verified by the study of photon number fluctuations of a single-mode laser at the threshold of lasing \cite{my}, confirming that it is a PT problem describable by GL theory \cite{hh}.  On the theoretical side, it has also been found that using the 2D Ising model to simulate quark-hadron PT the resulting scaling behavior of $F_q$ is in agreement with Eqs.\ (\ref{2.5}) and (\ref{2.6}) \cite {cgh}.  It does not mean, however, that an analysis for $F_q$ in the current data from heavy-ion collisions can verify or falsify the connection between hadronization and second-order PT because of the complications that are present in such systems but absent in the optical system.  The following sections are aimed at addressing such complications. 
We note that in the discussion above concerning Eqs.\ (\ref{2.1}) to (\ref{2.6}) no mention is made of the specifics about multiplicities, whether hadrons or quarks. In the remainder of this section we shall use particles as a generic term that can refer to either hadrons or quarks. The result to be obtained will be used in the following sections to generate configurations of quarks (and/or antiquarks) just before hadronization.

From a different perspective it is of interest to study the problem of clustering, since critical phenomenon is known to be characterized by clusters of all sizes.  It is a subject relevant to our investigation here because we shall analyze the 2D spatial distribution of particles in the $\eta$-$\phi$ space.  Cluster formation in the context of hadronization has been considered before \cite {hlp}, but from the point of view of self-organized criticality instead of second-order PT.  Since the latter exhibits the behavior of Eqs.\ (\ref{2.5}) and (\ref{2.6}), it is not difficult to see how such a behavior can be achieved by an adjustment of the scaling property of the cluster formation.

As a generic problem on clusters of particles, let $C$ be the number of particles in a cluster, and $P(C)$ be the probability  distribution in $C$, which we assume to have the scaling form
\begin{eqnarray}
P(C) \propto C^{-\gamma}, \quad \gamma > 0.     \label{2.7}
\end{eqnarray}
We let the center of a cluster be distributed randomly in a 2D space which we take to be 1 unit of length on each side.  We further let a particle in a cluster $C$ to be distributed randomly around its center with a Gaussian width of 
\begin{eqnarray}
\sigma = 0.1\ C^{-1/2}     \label{2.8}
\end{eqnarray}
so that there can be high density of particles, though with decreasing probability at high $C$.  Such spikes of multiplicity are what can give rise to intermittency.

To investigate the multiplicity fluctuations in small bin sizes, we divide the unit square into $M\times M$ bins, $M$ varying from 8 to 70.  We let $N_0$ particles be distributed in the unit square in accordance to Eq.\ (\ref{2.7}) and (\ref{2.8}), and allow $N_0$ to be approximately 100, since the sum of all clusters may not be exactly the same for every event.  With $n$ being the number of particles in a bin, we perform the calculation for $F_q(\delta)$ according to Eq.\ (\ref{2.1}), by averaging over all bins first, and then averaging over all $(5\times 10^5)$ events for any fixed $\delta = 1/M$.  Only bin multiplicities with $n \geq q$ are counted in $F_q(M)$.  The result for $\gamma = 2$ is shown in Fig.\ 1(a) that exhibits non-trivial dependence on $M$; it is not strictly linear in the log-log plot.  However, when $F_q$ is plotted against $F_2$ for $q=3, 4, 5$, very good linearity is found, as shown in Fig.\ 1(b) replicating the behavior found for GL \cite{hn}.  Similar power-law behavior is obtained for other values of $\gamma$.  We show in Fig.\ 1(c) the dependence of the exponent $\beta_q$, defined in Eq.\ (\ref{2.5}), on $q-1$, and find in accordance to Eq.\ (\ref{2.6}) that $\nu = 1.136, 1.171$, and $1.315$, for $\gamma = 1.5, 2.0, 2.5$, respectively.

\begin{figure}[tbph]
\centering
\includegraphics[width=0.5\textwidth]{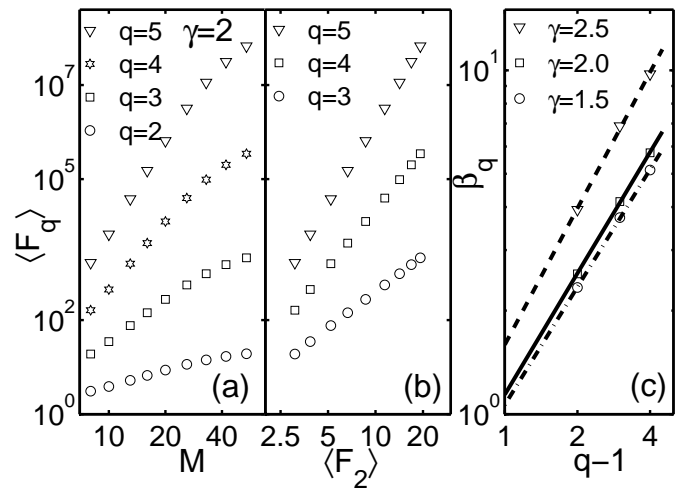}
\caption{Intermittency analysis of clustering model for $\gamma=2$ in Eq.\ (\ref{2.7}): (a) scaling in $M$, (b) $F$-scaling in $F_q$ vs $F_2$, and (c) power-law behavior of $\beta_q$ in Eq.\ (\ref{2.6}).}
\end{figure}

What we have done above is to find a quick way to relate a particle distribution in a 2D space to a GL-type $F_q$-scaling behavior through the use of a cluster distribution.  Since a range of values of $\nu$ has been obtained, the relationship between $\gamma$ and $\nu$ has no specific dynamical significance.  It merely demonstrates that large local multiplicity fluctuations can generate large $F_q$ with scaling properties.  In the case of $\gamma = 2.5$ we see in Fig.\ 1(c) that the straight line connecting the three points, when extrapolated, does not go through $\beta_2 = 1$, thus showing slight deviation from Eq.\ (\ref{2.6}).  The special value $\nu = 1.3$ for second-order PT can approximately be achieved by using a value of $\gamma$ between 2.0 and 2.5.  Although no dynamics has been introduced to establish that connection, our objective here is accomplished by having found a simple procedure to simulate configurations that may be relevant to a quark and antiquark system that is at the edge of PT to hadrons.  This is only the starting point of a much more complicated problem in heavy-ion collisions that we shall describe next.  

\section{Simulation of Events with Spatial Fluctuations in Heavy-ion Collisions}
If there is no dynamical structure in the spatial pattern of soft hadrons in the lego plot (of $\eta$ and $\phi$), then there should be nothing interesting in whatever measure that is used to analyze the data.  The reverse is, however, not necessarily true.  Unless a measure is sensitive to the consequence of certain dynamics, one may not observe what is interesting.  To provide motivation for an experimental effort to search for an unconventional signal, it is necessary to demonstrate the worthiness of such an effort under the best of circumstances.  The optimal scenario is that the quark-gluon system undergoes a second-order PT in forming hadrons so that there are large multiplicity fluctuations.  It is not known whether such a phenomenon occurs at LHC.  Our aim here is to simulate events that belong to different classes of dynamical characteristics, ranging from robust criticality to mundane randomness.  For each class of hadronization the measure to be used in the following section should exhibit distinguishing features in the hope that analysts of the real data would have the incentive to pursue the more difficult task of extracting worthwhile information from what is actually observed.

The physics we aim to simulate is for the duration between the end of the quark phase and the beginning of the hadron phase.  It is assumed that the density is so low that the confining forces among the quarks and the antiquarks begin to redistribute them spatially.  Gluons are assumed to have been converted to $q$ and $\bar q$ that are the basic units prior to hadronization.  We restrict our attention to the central rapidity region with $|\eta|  \leq 1$ and $0 \leq \phi \leq 2\pi$ for central collisions.  The surface of the plasma cylinder with those values of $\eta$ and $\phi$ is mapped to the unit square $S$.  
Since the fluctuation properties to be discussed below do not depend on the precise area that $S$ corresponds to, a portion of the $(\eta,\phi)$ region in the actual data should suffice to serve as a workable basis for analysis.
The cylinder hadronizes at the surface only, layer by layer.  Thus we have to consider many time steps $t_i, i=1, 2, \cdots$.  Between adjacent time steps the $q$ and $\bar q$ adjust their positions in $S$ in accordance to a procedure that we impose to simulate confinement and pion emission.  The quarks have thermal $p_T$ distribution whose inverse slope $T_i$ decreases incrementally with $t_i$.  At a new time step a new set of $q$ and $\bar q$ are introduced to represent the movement of the next layer of quarks to the surface.  The confinement procedure starts over again until most of the quarks (not necessarily all) are pionized before another time step is taken.  This is a general outline of the algorithm to convert $q$ and $\bar q$ to pions, whose coordinates in $(p_T, \eta, \phi)$ are registered for event-by-event analysis later.

The main part of our physics input is the deconfinement to confinement transition between time steps.  The principal characteristic of a critical phenomenon is the tension between the ordered and the disordered motions of a system.  In the Ising model of a magnetic system the near-neighbor interaction between spins tends to align all spins in the same direction, but the thermal motion tends to randomize them.  If the quark-hadron PT is in the same class of criticality, then the tension is between confinement that draws the $q\bar q$ into an ordered pair and the disordered thermal agitation that tends to keep the quark system in the deconfined state.  On the surface of the plasma cylinder the $q$ and $\bar q$ may start to cluster as the density and temperature get close to the transition point, thereupon a local region of high density contracts under the confinement force to improve the likelihood of a $q\bar q$ pair to fall within a confinement distance.  In principle, we should keep track of the colors of the quarks, but that would complicate the procedure much more.  The simple algorithm we adopt mimics the general idea of color confinement and generates large multiplicity fluctuations.  One of the chief objectives of this work is to see whether many such local fluctuations at different time steps can survive the superposition at the end of the hadronization process, and still be detected by an effective measure.  The feasibility issue is addressed at the cost of precision in QCD, which is difficult in the soft regime over an extended period.

Let us describe first the case of critical hadronization, followed later by other cases that are less critical.

\subsection {Critical}
\subsubsection{Initial Configuration}
We start by seeding the unit square $S$ with 500 pairs of $q \bar q$.  They are clustered according to the probability distribution $P(C)$, given in Eq.\ (\ref{2.7}).  That is, $C$ pairs of $q \bar q$ are placed in a cluster centered at a random point in $S$; they are Gaussian distributed around the center with a width specified by Eq.\ (\ref{2.8}).  As a result of the study in Sec. II that relates the exponents $\gamma$ and $\nu$, we choose $\gamma = 2$, corresponding to $\nu$ not quite up to 1.3.  As the cluster multiplicity $C$ is summed over all clusters, we stop the seeding process when the total just exceeds 500.  Inside each cluster the $q$ and $\bar q$ are not correlated; they are independently distributed in $(\eta, \phi)$ as well as in $p_T$, for which the thermal distribution is the exponential, $\exp(-p_T/T)$, with $T$ set at 0.4 GeV.

The above procedure is for setting up an initial configuration at $t_0 = 0$, counting from just before the first pions are emitted, but long after the collision time.  The value of $T$ is not set at $T_c$ which is lower, but at a value above the observed average $\left<p_T\right>$, since the bulk of the pions will be produced at later time when $T$ will be lower.  The clustering is put in to generate a spatial configuration that is most likely to represent a system moving toward a critical transition.

\subsubsection{Pionization}

In the above configuration if a $q$ and $\bar q$ are within a distance $d$ from each other that is less than $d_0 = 0.03$, then we regard the pair as effectively a pion and let it be taken away from the configuration, but registering the pion position in a separate $(\eta, \phi)$ space at the midpoint between the pair.  We assign a value of $p_T$ to that pion that is equal to the sum of the $p_{iT}$ values of the $q$ and $\bar q$.  This is based on the recombination model, where the recombination function has a momentum-conserving $\delta$-function: $\delta(p_T-p_{1T} - p_{2T})$ \cite{hy, gkl, rjf}. The thermal \dis s of quarks and pions have the same $T$. We ignore the color and flavor of the quarks without losing the essence of local fluctuation, since if the $q$ and $\bar q$ had color and flavor labels, it would take longer in the iteration process for the $q \bar q$ pair in a cluster to pionize without changing the $(\eta, \phi)$ coordinates appreciably.

\subsubsection{Contraction}

The probability that a $q \bar q$ pair is within $d_0$ apart is small, so the majority of the quarks remain in $S$.  They are under the influence of the color forces to move toward confinement.  To describe that movement in a collective way rather than in terms of pair-wise interaction that is unrealistic, we adopt the contraction procedure as follows.

Let $S$ be divided into $5\times 5$ bins.  Calculate the number of $q$ and $\bar q$ in each bin.  Separate the bins into two types:  dense bins have more $q$ and $\bar q$ than the average bin multiplicity, and the dilute bins have less.  The difference between $q$ and $\bar q$ is ignored here.  If adjacent dense bins share a common side, they are grouped together as members of a cluster of dense bins.  Let $D$ refer to such a cluster of dense bins, which may have an irregular shape, but are connected.  Let $N_D$ denote the number of bins in $D$.  Define $\vec r_D$ to be the coordinates in $S$ that is the center of mass of $D$.  Now, we do a contraction of $D$.  That is, we redistribute all the $q$ and $\bar q$ in $D$, centered at $\vec r_D$, but with a Gaussian width
\begin{eqnarray}
\sigma_D = \sigma_1N_D^{1/2},     \label{3.1}
\end{eqnarray}
where $\sigma_1$ is a parameter that characterizes the degree of contraction.  We use $\sigma_1 = 0.1$ here with other possibilities to be discussed later.  Because of the Gaussian distribution, the $q$ and $\bar q$ that are spread out originally in $N_D$ bins are drawn closer together to be located  mostly within the Gaussian peak, resulting in a contraction.  Since there are $5\times 5$ bins, the original bin size is $0.2\times 0.2$, to which $\sigma_D$ should be compared after contraction.
This is how we model the effect of the ordered motion due to confinement.  We repeat step {\it 2} to allow $q \bar q$ pairs to pionize in this new configuration.

\subsubsection{Randomization}

The disordered motion that counter-balances the ordered motion is the thermal randomization.  We model that by requiring all the $q$ and $\bar q$ in the dilute bins to be redistributed randomly throughout $S$, resulting in a new configuration.  We then repeat step {\it 3} to have another round of contraction and pionization.  
Each time in the iteration process more $q \bar q$ pairs are converted to pions, as we alternate contraction and randomization until around 95\% of the $q \bar q$ system is depleted.  We regard that as the end of one time step in which the quarks on the cylinder surface are hadronized.  We then proceed to the next time step when a new layer of quarks move up to the surface. 

We note that before the next time step is taken, pionization becomes increasingly difficult when fewer and fewer \qaq\ remain to find their partners to coalesce. Decreasing $\sigma_1$ to 0.05 speeds up the process near the end, but does not change the pion \dis\ in \ep\ very much. Physically, it is not necessary that all \qqb\ in a layer  hadronize before the next layer of quarks moves up. Leaving roughly 5\% to be mixed with the next layer of \qqb\ seems reasonable in our attempt to model a continuous process of hadronization by discrete steps.

\subsubsection{Subsequent Time Steps}

To advance to the next time step we introduce 100 new pairs of \qqb\ according to the \dis\ $P(C)$ and add them to the existing \qaq\ that remain from the previous step. We then follow the same procedure as above to contract, pionize, and randomize repeatedly until 5\% of \qaq\ remains. At the $i$th time step all $q$ and $\bar q$ have the \pt\ \dis\ with an inverse slope
\bq
T_i=T_{i-1}-0.02\ {\rm GeV}.  \label{3.2}
\eq
That is the $T_i$ that the pions emitted at $t_i$ will also have.
This is carried out 10 times so that in total we introduce 1500 \qqb\ paris, most of which turn into pions. To have approximately 1400 pions produced in $|\eta|<1$ in central collision at LHC is not unrealistic. Since we are not interested in global multiplicity fluctuations in this study, we stop after ten steps. The pions collected in $(p_T,\eta,\phi)$ for the event is then stored for later analysis of local fluctuations.

\subsection{Quasi-critical}

Consider now the case where clustering does not occur in the initial configuration, nor when new layers of \qaq\ move to the surface. Thus the initial 500 pairs of \qqb\ are seeded randomly in $S$, and so are the subsequent 100 pairs at each time step. There is then no correspondence with the second-order PT discussed in Sec.\ II. Technically, this corresponds to the exponent $\gamma$ in Eq.\ (\ref{2.7}) being very large, because the probability for having a large cluster $C$ is then very small. Specifically, we choose $\gamma=5$ for no clustering.
One may regard this case as being in correspondence to a cross-over in the phase diagram where no distinct boundary between the quark and hadron phases can be identified. However, we require that the confinement forces to be still at work to turn quarks to hadrons. So we use the contraction-randomization procedure described in subsection A to carry out pionization. Such a procedure simulates the tension between confinement and deconfinement, but cannot be regarded as what is needed for a critical transition.
The coordinates of each pion in $(p_T,\eta,\phi)$ are recorded as before for later analysis.

\subsection{Pseudo-critical}

Suppose now that we seed the configurations with clustering, but do not impose contraction between time steps. Thus the \qaq\ configurations, as each layer reaches the surface, are close to the critical condition, when we set $\gamma=2$ as we have done above. However, the hadronization process is carried out by letting \qqb\ pairs form pions whenever a pair gets close together. Without contraction we cannot require the distance between pairs to be less than $d_0$ as in Subsec.\ A.2, since the probability for that to occur is low. Without new dynamics during hadronization we simply let each quark to search for the nearest antiquark within a distance $d=0.1$ and let a pion be formed at the position midpoint between the \qqb\ pair. Then we go to another quark and repeat the process. When no more pairs can be found within that distance, we proceed to the next time step whatever numbers of \qaq\ are left. With more \qqb\ pairs supplied from the next layer, the probability of pionization is increased. This process will not convert all \qqb\ pairs to pions since the quarks can be far apart from antiquarks without contraction. To have quarks left over at the end of ten time steps does not matter as far as the local multiplicity fluctuation is concerned. We shall see that there is still interesting structure that can be extracted from the accumulated events simulated that way.

\subsection{Non-critical}

To the other extreme situation away from the above three cases we consider the non-critical case of  no organized dynamics at all, i.e., random configurations and  no contraction. The result should not have any content of interest. We have nevertheless carried out the simulation and will show the result that is non-trivial and therefore instructive.

\vspace*{.5cm}

A summary of the four cases can be expressed as a matrix shown below.
\begin{widetext}
\begin{equation}
\begin{array}{|c||c|c|c|}\hline
{} &\quad {\rm clustering} &\quad {\rm no\ clustering}\quad \\ \hline\hline
\quad{\rm contraction} \quad&\quad {\rm\bf critical}\quad&\quad {\rm\bf quasi{\rm-}critical}\quad \\  \hline
\quad{\rm no\ contraction} \quad&\quad {\rm\bf pseudo{\rm-}critical}\quad &\quad {\rm\bf non{\rm-}critical} \quad\\ \hline
\end{array}   \label{3.3}
\end{equation}
\end{widetext}

\section{Moments for Event-by-event Fluctuations}

We now consider the method of analysis of the many events either as measured at LHC or as generated in the models described in the preceding section. Because of the high multiplicity of particles produced in central collisions, we make cuts in \pt, such as in an internal $\Delta p_T$ around $p_T=1$ GeV/c. We have considered $\Delta p_T=0.04, 0.07$ and 0.1 GeV/c. In such small intervals the particle multiplicities are significantly reduced and spatial patterns in $(\eta,\phi)$ begin to appear because of the possibility of empty bins. 

To find an effective measure of the fluctuations we need it to be sensitive to both the spatial variations from bin to bin and the event-by-event fluctuations. We refer to the former as horizontal and the latter as vertical. Different moments are to be taken for horizontal and vertical averages so as to allow spatial and event-wise fluctuations to manifest separately.

We divide the unit square into $M^2$ bins with $M$ being not more than 70, depending on the multiplicity in the $\Delta p_T$ interval, just so that the important part of the $M$ dependence is captured. For the spatial fluctuations  we use the horizontal factorial moments for  event $e$, defined as 
\bq
F^e_q(M)=f^e_q(M)/[f^e_1(M)]^q ,  \label{4.1}
\eq
where
\bq
f^e_q(M)=\left< n(n-1)\cdots(n-q+1)\right>_h .  \label{4.2}
\eq
The average $\left< {\ }\right>_h$ is performed over all $M^2$ bins, $n$ being the multiplicity in a bin. Only $n\ge q$ is counted in $f^e_q(M)$. Clearly, if $M$ is very large, $f^e_q(M)$ may be zero, for $q\ge 2$, although $f^e_1(M)$ is never zero. However, there may be an event where $F^e_q(M)$ may not vanish at a large $M$; then it would imply sharp spikes of multiplicity in some bins. If the fluctuations among the bins are Poissonian, then we have $F^e_q(M)=1$ for any $M$, as is the case with vertical fluctuations discussed in Sec. II. Interesting spatial fluctuations are, however, not Poissonian.

For each event we can calculate $F^e_q(M)$. We note that $F^e_q(M)$ is a simple characterization of spatial pattern, but is not the only one possible. Any alternative description can also be used in the study below on event-by-event fluctuations. One may therefore  regard \fq\ as a generic symbol. 

If $\left<F_q(M)\right>_v$ denotes the vertical average of $F^e_q(M)$ over all events, then the fluctuation of $F^e_q(M)$ from $\left<F_q(M)\right>_v$ is what we want to quantify. To that end we consider the $p$th-order moments of
\bq
\Phi_q(M)=F^e_q(M)/\left<F_q(M)\right>_v ,  \label{4.3}
\eq
i.e.,
\bq
C_{p,q}(M)=\left<\Phi_q^p(M)\right>_v ,  \label{4.4}
\eq
which is a double moment introduced earlier for the study of chaotic behavior of particle production in branching processes \cite{ch}. Whereas $q$ must be an integer, $p$ need not be. In fact, the derivative of $C_{p,q}(M)$ at $p=1$, i.e.,
\bq
\Sigma_q(M)=\left.{d\over dp}C_{p,q}(M)\right|_{p=1} = \left<\Phi_q \ln \Phi_q\right>_v , \label{4.5}
\eq
has been related to an entropy defined in the event space \cite{ch}, but by itself it is not very useful because it depends on $M$.
Simplification can occur  if $C_{p,q}(M)$ has a power-law behavior in $M$
\bq
C_{p,q}(M) \propto M^{\psi_q(p)} .  \label{4.6}
\eq
Then an (entropy) index can be defined as
\bq
\mu^{(1)}_q={d\Sigma_q(M)\over d \ln M} = \left.{d\over dp}\psi_q(p)\right|_{p=1} , \label{4.7}
\eq
which is independent of $M$. It was found that $\mu^{(1)}_q$ can characterize the fluctuations of spatial patterns so well that they are as useful as the Lyapunov exponents in providing a quantitative measure of classical chaos \cite{ch1}. 

 The power-law behavior of $C_{p,q}(M)$ in Eq.\ (\ref{4.6}) has been referred to as erraticity \cite{h1,ch2}.  It was proposed as 
 the next logical step to take beyond the intermittency analysis. Attempts
 have been made to find   experimentally the erratic fluctuations of $F^e_q$ from event to event in multiparticle production \cite{kd}.  In meson-proton collisions at 250 GeV/c beam momentum, it was found that the erracticity measures are dominated by statistical fluctuations at such low energy \cite{ata}. In nucleus-nucleus collisions interesting signals have been found both in models \cite{fml} and in emulsion experiments \cite{dg} even at low energies.
At high energy where we can examine the deconfinement-to-confinement transition, it  will become clear in the next section, where model calculations can generate concrete local multiplicity fluctuations, that it is more comprehensive to consider the range $1 \le p \le 2$ in the study of $C_{p,q}(M)$. If in that range $\psi_q(p)$ has a linear dependence on $p$, it is better to define the slope in that wider range as
\bq
\mu_q=d\psi_q(p)/dp .  \label{4.8}
\eq
It is this quantity $\mu_q$, referred to as erraticity indices, that will become an effective measure of the criticality classes that is independent of $p$ and $M$.
A large value of $\mu_q$ means that there are influential contributions to $C_{p,q}(M)$ from large $F^e_q(M)$ weighted more heavily at large $p$, which in turn implies that very erratic \fl s of the spatial patterns are involved to render $F^e_q(M)$ non-vanishing at large $q$ and $M$.

\begin{figure}[tbph]
\centering
\hspace*{-.5cm}
\includegraphics[width=.5\textwidth]{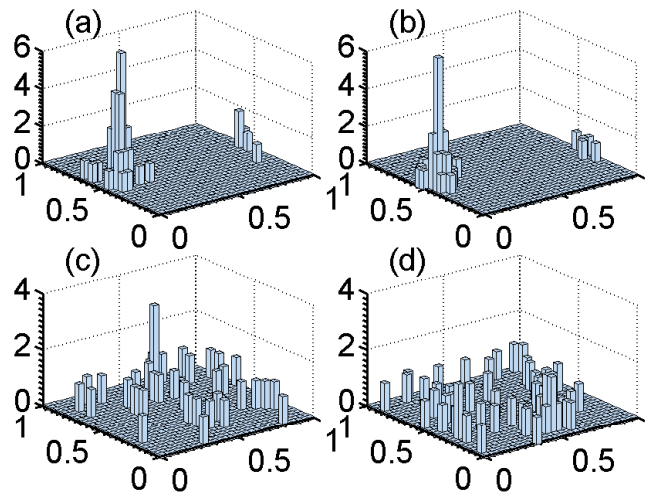}
\caption{Examples of bin multiplicity fluctuations in $(\eta,\phi)$ for the four cases arranged in the matrix form of (\ref{3.3}), i.e., (a) critical, (b) quasi-critical, (c) pseudo-critical, (d) non-critical.}
\end{figure}

\section{Results of Model Calculations}

We have simulated in the order of $10^6$ events for the four models ranging from critical to non-critical cases described in Sec.\ III. To give visual images of their qualitative differences, we select one event from each case to illustrate their behaviors in the lego plots. 
For an event to contribute to $F_q$ at $q\ge 2$, it cannot be typical if the average bin multiplicity $\left< n\right>$ is small. We show untypical events that have large $n$ in some bins and thus can contribute to non-trivial $F_3$ at $M=30$ and $\Delta p_T=0.1$ around $p_T=1$ GeV/c, for which $\left< n\right>$ is approximately 0.1.
In Fig.\ 2 the four lego plots are arranged as a matrix in the classification given in (\ref{3.3}).  
Each of the plots corresponds to an event in a given class that contribute to the largest value of $F_3(M)$ at $M=30$; thus there should be at least one bin that has a bin multiplicity $n
\ge 3$. It is evident that the distinctive difference among them is that the pseudo-critical and non-critical cases in the lower  panels (c) and (d) have particles distributed more widely throughout the base square than in the other two cases above in (a) and (b), where a localized cluster surrounds a high peak. The intent of this figure is only to give a sense of the qualitative difference in the nature of spatial patterns  when a stringent demand is placed on the bin multiplicity to depart from the average.

 \begin{figure}[tbph]
\centering
\includegraphics[width=.5\textwidth]{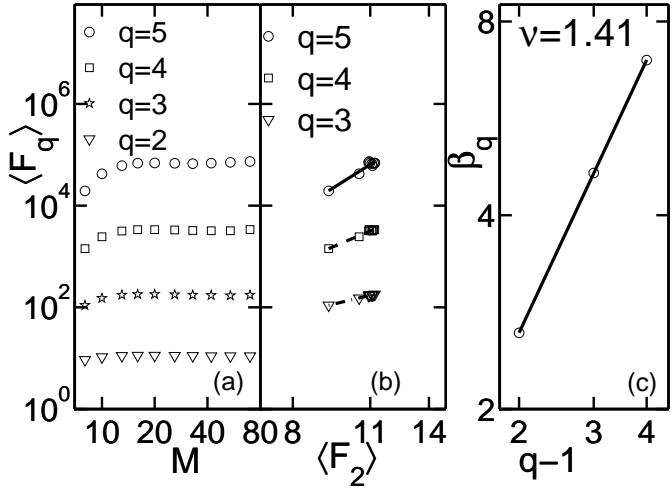}
\caption{Intermittency analysis for the critical case. Panels are the same as in Fig.\ 1.}
\end{figure}

\begin{figure}[tbph]
\centering
\includegraphics[width=.5\textwidth]{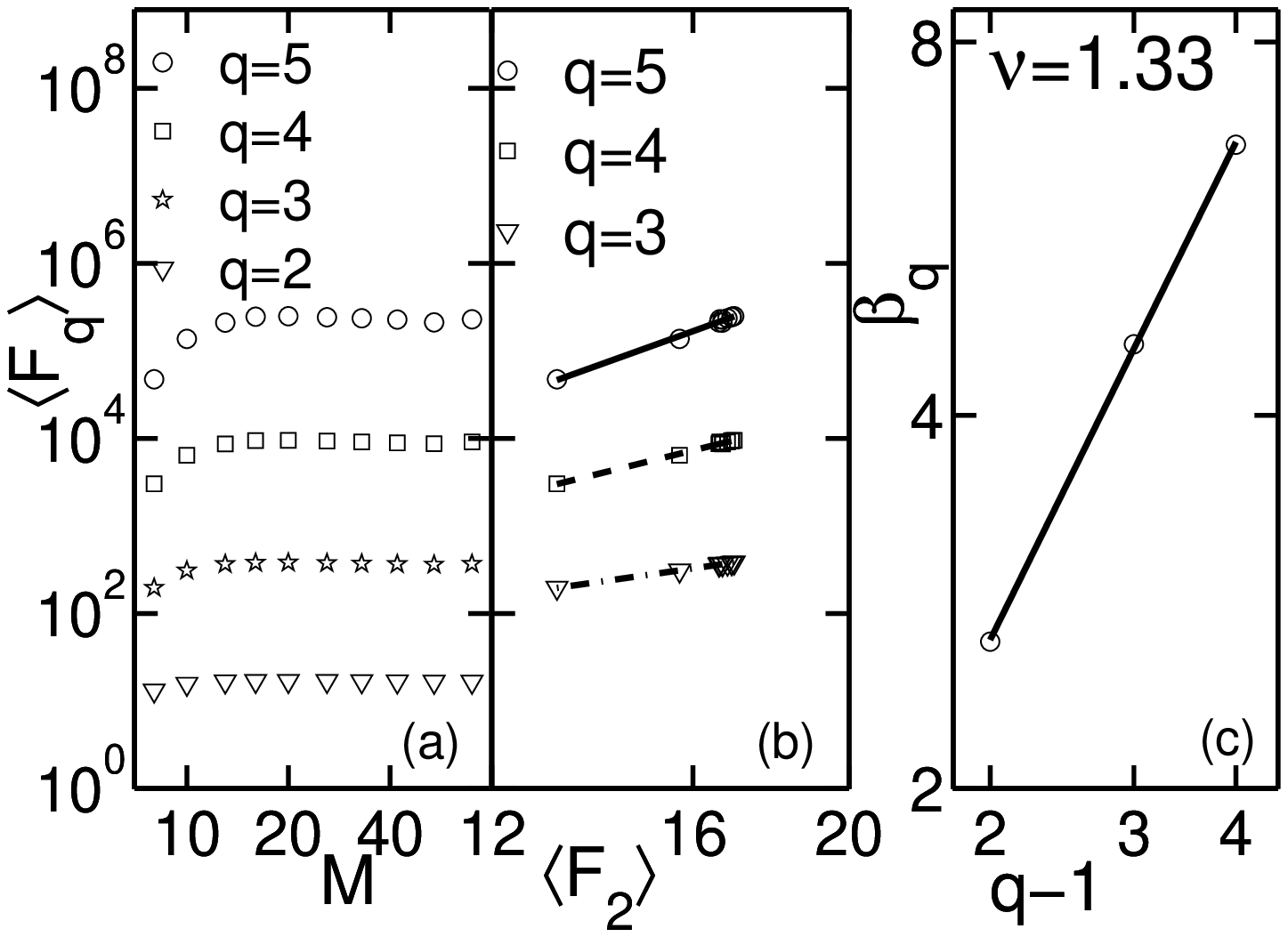}
\caption{Intermittency analysis for the quasi-critical case. Panels are the same as in Fig.\ 3.}
\end{figure}

\begin{figure}[tbph]
\centering
\includegraphics[width=.5\textwidth]{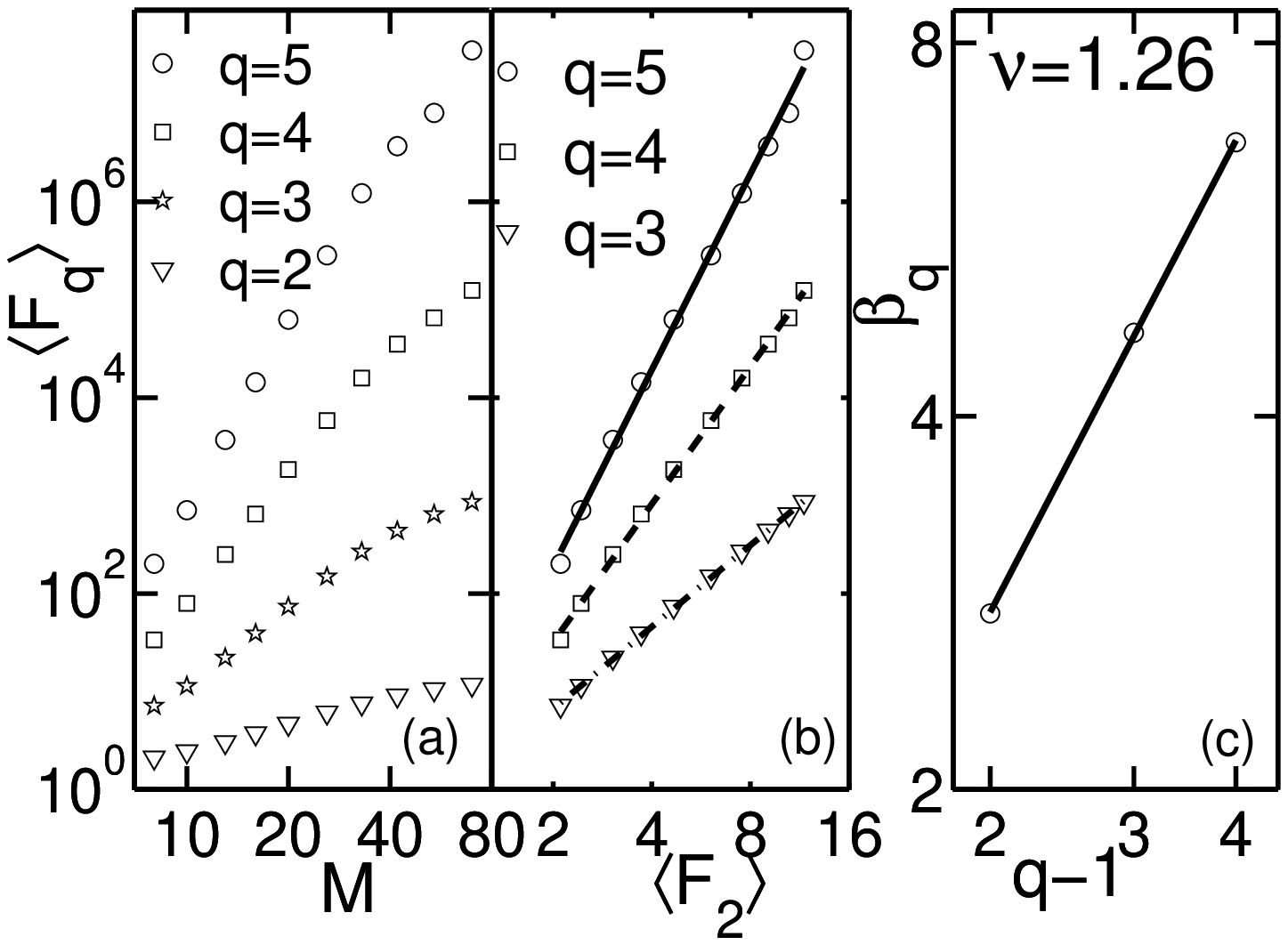}
\caption{Intermittency analysis for the pseudo-critical case. Panels are the same as in Fig.\ 3.}
\end{figure}

\begin{figure}[tbph]
\centering
\includegraphics[width=.5\textwidth]{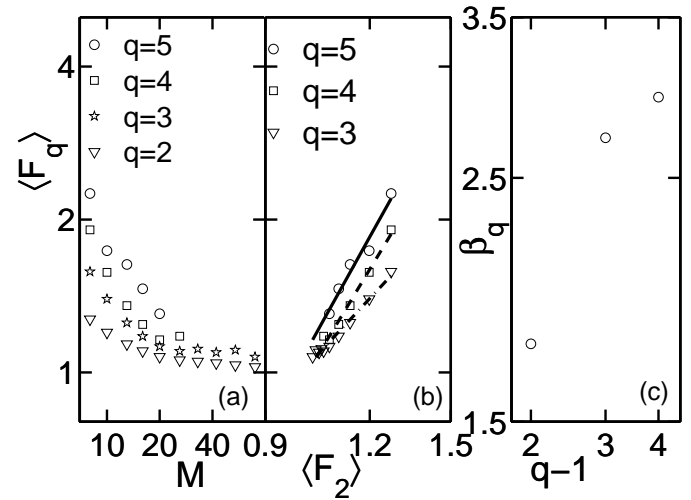}
\caption{Intermittency analysis for the non-critical case. Panels are the same as in Fig.\ 3.}
\end{figure}

To see the scaling behavior, we examine
 the vertical averages $\left<F_q(M)\right>_v$ vs $M$ in log-log plots for $q=2, \cdots, 5$ and for the cut $\Delta p_T=0.1$ GeV/c. Hereafter the subscript $v$ will be omitted for brevity.
 For the critical case the results are shown in Fig.\ 3 where in (a) there is an increase with $M$ before $M$ gets larger than 20, but they all have similar behavior for different $q$, so when $\left<F_q\right>$ is plotted against $\left<F_2\right>$, we see in (b) a simple straight-line behavior for $q=3, 4, 5$. Using Eqs.\ (\ref{2.5}) and (\ref{2.6}) to describe the power $\beta_q$, we find in (c) that $\nu_{\rm crit}=1.41$. Similar properties are found in the quasi-critical case with $\nu_{\rm quasi}=1.33$, as shown in Fig.\ 4. In the pseudo-critical cases, shown in Fig.\ 5, we see robust scaling behavior in (a). $F$-scaling behavior in (b) yields the value $\nu_{\rm pseudo}=1.26$ shown in (c). Finally, in the non-scaling case we find the opposite situation where $\left<F_q(M)\right>$ decreases with increasing $M$, as exhibited in Fig.\ 6 (a). It means that bin multiplicities do not get large enough deviations from $\left< n\right>$ through random fluctuations so that, when the bin size gets small, $\left<F_q(M)\right>$ approaches 1 that one expects from Poissonian fluctuations, as stated just below Eq.\ (\ref{2.4}). In (b) we still see regularity in $\left<F_q\right>$ plotted against $\left<F_2\right>$, but it is important to recognize that the low end corresponds to high $M$ with all $\left<F_q\right>$ around 1, while the high end is for low $M$, quite contrary to the three cases in Figs.\ 3-5. One can extract the values of $\beta(q)$ as in (c), but there is no sensible value of $\nu$ to be assigned to this case. In short, the non-critical case does not yield any interesting result from the study of this kind. We have to go beyond simple intermittency analysis in order to find a suitable description that can render quantitative comparison between the critical and non-critical cases.
 
  Although the results shown in Figs.\ 3-5 are clear and easily quantifiable in terms of $\nu$, a great deal of  information is lost by calculating those averages.
 To exhibit the degree of fluctuations of the spatial patterns from the average, let us use $P(\Phi_q)$ to denote the probability \dis\ of event $\Phi_q(M)$   at fixed $M$, where $\Phi_q(M)$ is defined in Eq.\ (\ref{4.3}). In Fig.\ 7 we show $P(\Phi_q)$ for $q=2$ and 4, and for clarity only for $M=8$ and 30. The four classes of criticality are again in the matrix format of (\ref{3.3}). We see that in all cases the \dis s for $q=2$ and $M=8$ (solid lines) are peaked at $\Phi_q=1$. But for other values of $q$ and $M$, the four cases differ in different ways. In the other extreme situation corresponding to $q=4$ and $M=30$ (lines with crosses), we see that $P(\Phi_4)$ is peaked at $\Phi_4=0$ in all cases. That is because in small bins the average bin multiplicity $\left< n\right>$ is much less than 4, so the values of $\Phi_4$ for many events are 0. In fact, for (c) pseudo-critcial and (d) non-critical, only a small fraction of events have large enough bin fluctuations to render $F_4^e$ non-zero, so $P(\Phi_4)$ has a $\delta$ function peak at $\Phi_4 = 0$, whose positions are shifted in Fig.\ 7 for visibility's sake. In the intermediary values of $q=4, M=8$, the dashed lines in all four cases are all very broad, signifying wide fluctuations. For $q=2, M=30$ (dash-dotted lines) the peaks are around $\Phi_2=1$, similar to the solid lines.
 
 \begin{figure}[tbph]
\centering
\hspace*{.05cm}
\includegraphics[width=.5\textwidth]{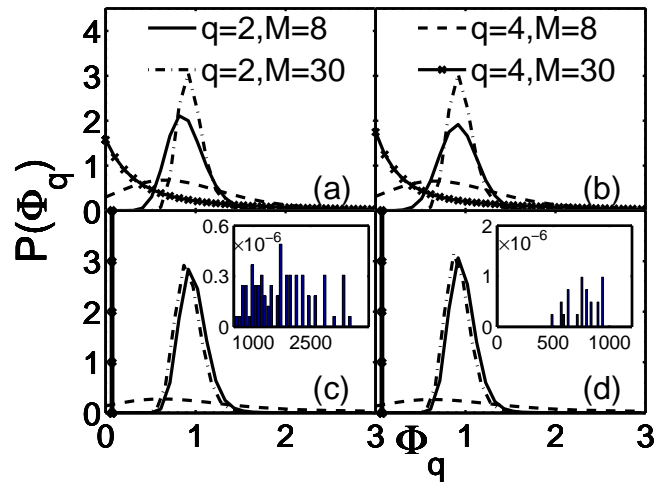}
\caption{Probability distributions of $\Phi_q(M)$ for various values of $q$ and $M$. The panels correspond to the criticality cases expressed in (3.3) and as shown in Fig.\ 2. The insets in (c) and (d) are for large values of $\Phi_q$ as explained in the text. The vertical axes have a common scale factor of $10^{-6}$.}
\end{figure}
  
 In the insets of Fig.\ 7 (c) and (d) we show that  for $q=4$ and $M=30$ there are contributions to $P(\Phi_4)$ at extremely large $\Phi_4$ (in the order of $10^3$). 
 The vertical axes have the scale factor $10^{-6}$. They balance the $\delta$ functions at $\Phi_4=0$ so that the average is $\left<\Phi_4\right>_v=1$, by definition. This irregular behavior reveals the nature of fluctuations from event to event. When the average bin multiplicity $\left< n\right>$ is about 0.03, it is difficult to find events in which there is a bin with $n\ge 4$, unless there are dynamical effects (such as confinement contraction) to introduce large fluctuations. In cases (c) and (d) almost all events have $n<4$ so $\Phi_4=0$, except for some very rare events that make non-trivial contribution to non-zero $\Phi_4$, whose values are therefore very large because $F_4(M=30)$ is exceedingly large (due to the smallness of $f_1^4$) even though $\Phi_4$ is normalized by $\left<F_4(M=30)\right>$ which is proportional to the rarity of such events.
 
The probability \dis s $P(\Phi_q)$ contain too much information that cannot easily be conveyed. 
 We learn from Fig.\ 7 that for $q=2$ there are no drastic differences among the four cases when $M$ is increased from 8 to 30. It means that bin multiplicities in each case can fluctuate sufficiently to exceed $n=2$ and generate a modest width of the peaks in $P(\Phi_2)$ around $\Phi_2=1$. We therefore should not expect a good measure at $q=2$ to distinguish the criticality classes. For $q=4$, however, we see significant differences between the cases (a,b) with contraction and (c,d) without contraction. To quantify their differences we consider the moments $C_{p,q}(M)$ defined in Eq.\ (\ref{4.4}). For $p=2, q=4$, we show in Fig.\ 8 the $M$ dependence for three intervals of $\Delta p_T: 0.04, 0.07$ and 0.1 GeV/c. The cases a, b, c, and d in the legend correspond to the panels in Fig.\ 7 arranged in the matrix form of   (\ref{3.3}). Some of the open symbols are displaced slightly from $M=8, 16$ and 30 in order to avoid them from  being covered by the filled symbols. We see that they satisfy power-law behavior very well in all cases, validating the meaningfulness of the erraticity exponents
 $\psi_q(p)$ defined in Eq.\ (\ref{4.6}). Similar study can be done for $p=1.25, 1.5$ and 1.75, and yield similar scaling behavior, which we do not show for brevity. 
 
 \begin{figure}[tbph]
\centering
\hspace*{.1cm}
\includegraphics[width=.5\textwidth]{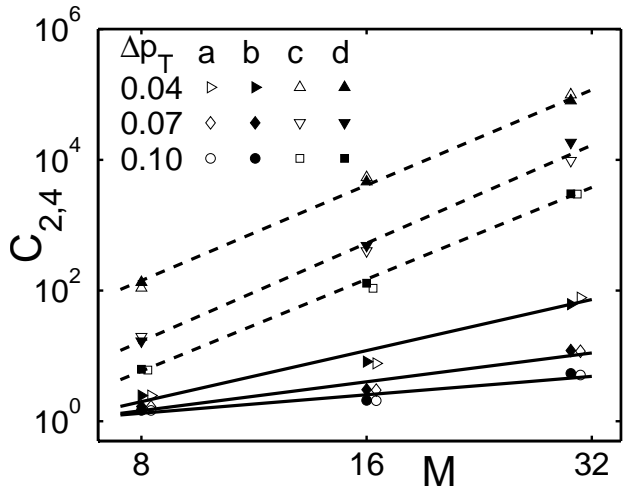}
\caption{Power-law behavior of $C_{2,4}(M)$ exhibiting erraticity. The labels a, b, c, d correspond to the four panels in Figs.\ 2 and 7 in the matrix format of (\ref{3.3}).  Pairs of symbols are approximated by one straight line, so 12 types of symbols are represented by 6 lines. $\Delta p_T$ are in the units of GeV/c and are around the value $p_T=1$ GeV/c.}
\end{figure}

\begin{figure}[tbph]
\centering
\includegraphics[width=.5\textwidth]{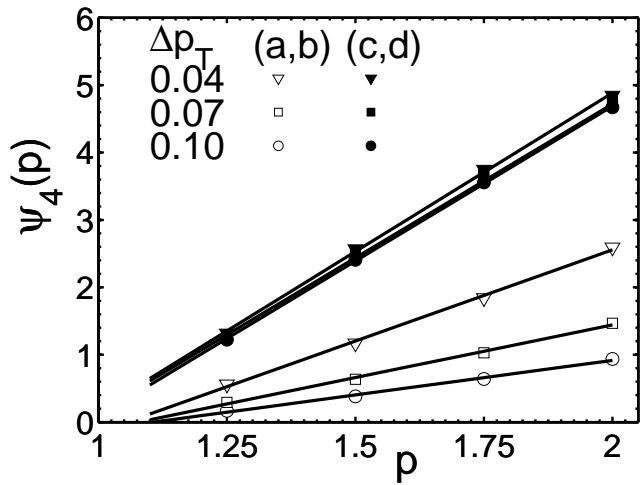}
\caption{Linear dependencies of the erraticity exponents $\psi_4(p)$ on $p$ for the (a,b) cases in open symbols and for the (c,d) cases in the solid symbols. Their slopes give the values of the indices $\mu_4$. The units of $\Delta p_T$ are in GeV/c.}
\end{figure}
 
 The values of  $\psi_4(p)$ are shown in Fig.\ 9. In that figure the cases (a) and (b) are depicted
collectively by open symbols, while the cases (c) and (d) are by filled symbols. The values
of  $\psi_4(p)$ in the (c,d) group are all larger than those in the (a,b) group. 
 Evidently, there are good linear dependencies on $p$ between 1.25 and 2 in all cases. 
The straight lines drawn through them are extended only to $p=1.1$. 
At $p=1$ it is necessary that $\psi_q(1)=0$ because $C_{1,q}(M)=1$ for any $q$ and $M$. The lines, if extrapolated further to $p=1$, would all miss the origin $(1,0)$ by little bits. There are some reason for that to happen at the point $p=1$, which we discuss below. For now, we concentrate on the linear portion in Fig.\ 9 and determine the slopes $\mu_4$, as  defined in Eq.\ (\ref{4.8}).
It is clear that the (c,d) group has essentially no dependence on $\Delta p_T$. Even in the (a,b) group the spread due to different $\Delta p_T$ cuts is not severe enough to make it unreasonable for us to give the three lines in each group an average slope as follows:
\bq
\mu^{a,b}_4 = 1.87 \pm 0.84 ,  \qquad  \mu^{c,d}_4 = 4.65\pm 0.06. \label{5.1}
\eq
It is remarkable that we can obtain numerical summary of the different cases independent
of $p$ and $M$ and only mildly dependent on $\Delta p_T$. The values of these indices mean that for the models (a,b) that have contraction due to
confinement there are spikes in bin multiplicity locally, making them easier to have bins
with $n \ge 4$ (thus lower values of $\mu_4$ or less erratic) than for the models (c,d) that have no contraction. In the latter cases there are
not enough multiplicity fluctuations to have $n\ge 4$ in most bins and in most events, so the
probability of having a non-zero $F_4$ at high $M$ is extremely low, but in those rare events the
value of $F_4$ is so high (thus more erratic) that the $p$th moment with $p \ge 1.25$ give them such a high weight as to
raise $\psi_4(p)$ and $\mu_4$ significantly above those for (a,b) cases.

Turning now to the complication at the point $p=1$, we note that if the straight lines in Fig.\ 9 go through the origin $(1,0)$ exactly, then the slopes are constant throughout $1\le p\le 2$ and we could alternatively focus on the point $p=1$ exclusively. In that case we could return to Eqs.\ (\ref{4.5}) and (\ref{4.7}) and determine $\mu_4^{(1)}$ directly from $\Sigma_4(M)$. Unfortunately, we have found that $\Sigma_4(M)$ does not depend on $\ell n M$ linearly, so we are unable to make use of the Eq.\ (\ref{4.7}) to calculate the slope at $p=1$. This is consistent with the fact that the  straight lines in Fig.\ 9 do not cross the origin precisely, and that some bending of those lines near the point $(1,0)$ reveals the lack of universality of the slopes for all $p$. Stated differently, the power-law behavior of Eq.\ (\ref{4.6}) is not true for all $p$; the local multiplicity \fl s can be so severe and are so different among the four cases that the indices $\mu_q^{(1)}$ at $p=1$ cannot summarize their differences.

Returning to the results expressed in Eq.\ (\ref{5.1}), we can conclude that what separates (a,b) from (c,d) is the dominating effect of contraction over clustering.
Recall from (\ref{3.3}) that the two columns are for cases where initial and reseeded configurations between time steps are  with clustering (a,c), and  without clustering (b,d), while the two rows are for cases where the configurations between time steps undergo contraction (a,b), and  no contraction (c,d). Since there are 10 substeps within each time step, contractions rearrange the configurations sufficiently so that the erraticity moments $C_{p,q}$ retain essentially no memory of the clustering effects in the input. Physically, it means that as the quark-gluon plasma approaches low density near the end of its expansion, whether or not there is critical clustering, the confinement forces that act on the quarks near the surface exert the dominant effect on drawing the $q\bar q$ pairs together in order to pionize, despite the opposing tendency to randomize due to thermal activities that persist. The same process is repeated each time a new layer moves to the surface. The tension between confinement and deconfinement is what leads to large local fluctuations evidenced by the low values of the erraticity indices $\mu_4^{\rm a,b}$. Without that tension the bin-multiplicity fluctuations have no dynamical push beyond randomness, so it is highly erratic to have some rare events to contribute to non-trivial $C_{p,4}$, hence higher values of $\mu_4^{\rm c,d}$.
  
  \section{Conclusion}
  
The purpose of this work is mainly to describe an unconventional method to analyze the LHC data in the hope that some experimentalists may find it adventuresome. The new frontier opened up by the high multiplicity events provides a fertile ground for exploration  that is not feasible at lower energies. That is why the subject is not among those predictions that could be extrapolated from RHIC. If there is any hint of critical behavior in the quark-hadron transition, that would be a new discovery at LHC. Even if nothing critical is found, analysis along the line suggested here should lead to deeper understanding of the hadronization process.

Our models of the four classes of criticality may not turn out to be realistic, but they have been useful in testing the effectiveness of the erraticity moments and indices. The basic issue is, of course, whether the proposed measures can be applied to the real data to uncover interesting physics. There exist other physical processes that are totally ignored in this study. Chief among them is minijet production, which is understood to be copious in Pb-Pb collisions at 2.76 TeV \cite{hz}. Usual jet study at LHC is for $p_T$ very large, e.g. $>50$ GeV/c. Our analysis here is for $p_T\approx 1$ GeV/c, which is minute compared to the jet towers, but the result can nevertheless be affected by thermal-shower recombination. It is unlikely that the low $p_T$ enhancement by minijets can lead to multiplicities greater than 4 in very small bins, but rare events with large fluctuations are what the erraticity moments $C_{p,q}(M)$ are sensitive to. Thus at this point we cannot rule out contamination of the phase transition effects by minijets. Such possibilities perhaps would encourage the experimentalists to investigate the subject, either to find resolution of ambiguities or to gain new perspective on an aspect of physics
 that is not well understood.

\section*{Acknowledgment}

This work was supported  in
part,  by the U.\ S.\ Department of Energy under Grant No. DE-FG02-96ER40972 and by the National Natural Science Foundation of China under Grant No.\ 11075061, and the Program of Introducing Talents of Discipline to Universities under Grant No.\ B08033.

\end{document}